\documentclass[aps,prapplied,reprint,superscriptaddress,amsmath,amssymb]{revtex4-2}

\usepackage{graphicx}
\usepackage[T1]{fontenc}
\usepackage{textcomp}
\usepackage{xcolor}
\usepackage{mathrsfs}
\usepackage{tabularx}
\usepackage{algorithmic} % Kept from your original setup

\begin{document}

\title{Simulation of exchange coupling effects in double quantum dot FinFET-like structures}

\author{Ilan Bouquet}
\email{bouqueti@iis.ee.ethz.ch}
\author{Alexander Maeder}
\author{Mathieu Luisier}
\affiliation{Integrated Systems Laboratory, ETH Zurich, Gloriastrasse 35, 8092 Zurich, Switzerland}

\begin{abstract}
By leveraging a GPU-accelerated Schrödinger-Poisson (SP) solver, we characterize exchange coupling in a hole spin double-qubit device involving a double quantum dot (DQD) system formed inside a 5-gate silicon fin field-effect transistor (FinFET) similar to real experimental structures. The self-consistent SP simulations rely on a finite difference discretization of the 3D volume and on a Luttinger-Kohn 6$\times$6 k$\cdot$p Hamiltonian accounting for magnetic fields and strain distribution. They return the gate-induced confined electronic states and the corresponding electrostatic potential hosting the DQD. These quantities serve as inputs to a two-particle Hamiltonian that is constructed from single-particle Slater determinants through the configuration interaction (CI) method. By diagonalizing this two-particle Hamiltonian, the eigenstates and eigenenergies of the DQD system are obtained, together with their exchange coupling. We show that our simulation framework, using a reduced number of basis states, is capable of reproducing the magneto-electrostatic behavior of the devices of interest, as predicted from theory and observed experimentally. We finally leverage our approach to determine the optimal operating conditions of a two-qubit quantum logic gate implemented in a Si FinFET structure.
\end{abstract}

\maketitle

\section{Introduction}

Silicon spin qubits take advantage of the highly mature complementary metal-oxide-semiconductor (CMOS) technology, making them one of the most promising approaches to quantum computing \cite{loss_quantum_1998,kloeffel2013prospects,burkard2023semiconductor}. The true advantage of Si qubits resides in the ability of the semiconductor industry to grow crystals and oxides with high quality and very low defect concentrations and to integrate billions of transistors onto the same chip. These features could pave the way for large-scale quantum processors capable of outperforming their classical counterparts on a range of important computational tasks, including integer factorization (Shor's algorithm \cite{shor1994algorithms}), unstructured database search (Grover's algorithm \cite{grover1996fast}), or the simulation of complex quantum systems \cite{feynman2018simulating}. Significant landmarks have already been achieved with electron quantum dots (QDs) for which single- and double-qubit gate fidelity exceeding 99\% have been reported using isotopically enriched planar Si \cite{yoneda2018quantum,mills2022two,noiri2022fast,borsoi2024shared,petit2020universal,philips2022universal}. Recent studies from the industry and targeting similar structures have demonstrated high-fidelity, multiple-QD electron spin qubit chips fabricated on standard Si-CMOS lines of production, with a planar configuration \cite{doi:10.1021/acs.nanolett.4c05205,petkov2025field,steinacker2025industry}. This milestone reaffirms the potential of the CMOS technology to massively scale up quantum architectures based on Si spin qubits. 

Despite their scalability potential, planar spin qubits based on electron QDs face fundamental physical limits. First, their weak intrinsic spin-orbit coupling (SOC) requires the integration of bulky external micromagnets, which increase the design complexity. Moreover, due to vertically arranged material stacks, the 2D electron gas formed within the semiconductor layer tends to spread out across a wide interface with typical radii between 20 and 50~nm \cite{veldhorst2015two}. As such, the area occupied by the qubit wavefunction is more likely to interact with interfacial inhomogeneities like charge traps or surface roughness, making it more susceptible to decoherence through charge-noise \cite{connors2022charge}. Also, because of the planar configuration of the channel, the gate-mediated electric fields only propagate vertically through the stack. As a consequence, a complex layout of multiple metallic gates is necessary to create the QD, thus limiting the ``sweet-spot'' tunability of fabricated devices. 

To address these shortcomings, attention has been increasingly turned to spin qubits hosted within hole QDs, as they benefit from the intrinsic spin-orbit coupling (SOC) present in the valence band of Si \cite{winkler_spin-orbit_2003}. The latter can be harnessed to induce spin-flip through fully electrical control, an effect known as electric-dipole-spin-resonance, thus reducing the number of components to be integrated altogether. To circumvent the aforementioned limitations of vertically layered stacks, theoretical works suggested to adopt non-planar semiconductor channels, as provided by three-dimensional nanowires and fin field-effect transistors (FinFET) \cite{kloeffel2018direct,milivojevic2021electrical}. On the one hand, a gate-all-around (GAA) architecture enhances electrostatic control and facilitates the formation of QDs, while diminishing cross-talk between adjacent QDs. On the other hand, carefully engineered channel geometries (e.g., core-shell nanowire, hut wire, fins) naturally confine QDs, leading to fewer number of physical gates than in two-dimensional devices to generate, couple, and electrically drive qubits \cite{ciriano2021spin,piot2022single,eggli2024all,kuhlmann2018ambipolar}. Another crucial advantage of three-dimensional structures stems from the increased heavy- and light-hole mixing they exhibit, which is known to boost SOC and generate a strong directional dependence in the magnetic response of the hole spin qubit \cite{kloeffel2011strong,froning2018single,adelsberger2022enhanced}. This mixing renders $g$-factors more anisotropic, allowing for greater flexibility in terms of magnetic field orientation, as compared to planar heterostructures, and for less constrained design guidelines. Intrinsic confinement and band mixing are indeed keys to the realization of fast-switching and, if properly operated, high-fidelity quantum gates. As an example, single-qubit gate fidelity with a fault-tolerance threshold of 99\% has been demonstrated in non-planar experimental devices based on both Si and Ge \cite{camenzind2022hole,froning2021ultrafast}. 

However, moving beyond single-qubit devices and fabricating highly reliable two-qubit quantum gates still remains a challenge. The coupling between two adjacent QDs and the hole charges they host gives rise to strong Coulomb interactions and complex exchange mechanisms between them \cite{loss_quantum_1998,burkard_coupled_1999}. The working principle of most two-qubit gates articulates itself around this exchange coupling interaction, which is responsible for coherent oscillations between the two-particle eigenstates constituting the core of the DQD system. In particular, the magnitude of this effect can be more efficiently controlled in 3D architectures with a GAA configuration due to their enhanced electrostatics, whereas entanglement can be achieved through precise pulsing of the gate coupling the qubits. This action leads to a lowering of the potential barrier separating the two dots, thus switching on the exchange interaction between them. Combined with single qubit rotation, the created two-qubit entangling gates serve as indispensable building blocks of universal quantum gate sets \cite{divincenzo1995two,deutsch1995universality,barenco1995elementary}. 

Precisely controlling the exchange coupling interaction mechanism in DQDs is therefore a fundamental requirement of quantum computation. In non-planar, multi-gate structures, this involves tuning several parameters (gate biases, external magnetic fields) whose individual impact and combined influence on the overall performance might be difficult to interpret. This is where the application of advanced numerical methods to model quantum systems can be really useful. By allowing to rapidly test different parameter configurations and by helping identify ``sweet spots'', device simulation can support on-going experimental efforts and the design of next-generation, multi-qubit quantum gates. Focusing solely on the exchange coupling, the Fermi–Hubbard model \cite{hubbard1963electron} provides an estimate of the interaction between two holes and its parameters can be fitted to reproduce experimental measurements \cite{jirovec2021singlet,saez2025exchange} or simulation results of higher physical complexity \cite{stepanenko2012singlet,abadillo2021two}. An example of more advanced, but computationally more intensive model is configuration interaction (CI), which can be used to rigorously describe the exchange coupling mechanism between adjacent dots \cite{sherrill1999configuration,szabo2012modern}. In the CI method, a two-particle interacting Hamiltonian is created by expanding all its entries in terms of a truncated basis of single-particle pairs obtained from the diagonalization of the non-interacting system. By leveraging this approach, the exchange coupling between two hole spin qubits was computed for a planar Ge device with less than 100 pairs \cite{rodriguez2025dressed}.

In this work, we focus on the calculation of the exchange coupling interaction between hole QDs confined inside a Si FinFET channel. As testbed, we use the geometry of an experimental 5-gate, $p$-type, Si-based, triangular FinFET \cite{geyer2024anisotropic} for which experimental measurements were conducted.
To investigate such structures, we have developed a quantum mechanical approach capable of revealing their functionality and of assessing their potential as suitable double-qubit quantum gates. Our GPU-optimized solver relies on self-consistent Schr\"odinger-Poisson (SP) simulations based on 6-band k$\cdot$p and including the strain response of the Si channel as it is cooled down to cryogenic temperatures. The resulting eigenstates are collected to form a reduced basis that, by applying the CI method, allows for rapid explorations of key figures-of-merit, e.g., exchange coupling constant or anti-crossing region, over a large voltage and magnetic field range. Comparison with experimental data is provided as well.
 
The paper is organized as follows: In Section \ref{sec:Methodology}, we present the FinFET geometry of interest and introduce the simulation framework we implemented to determine the exchange coupling in a gate-induced DQD system. In Section \ref{sec:Results}, we discuss the most relevant results obtained with our quantum mechanical model, focusing on the knobs that can be tuned through electrical and magnetic fields to ``program'' the device under test. We further highlight its optimal operating conditions to perform as $CZ$ and $SWAP$ two-qubit gate before comparing our numerical findings to real experimental measurements. Conclusions are drawn in Section \ref{sec:Conclusion}.

\begin{figure}[t]
\includegraphics[width=\columnwidth]{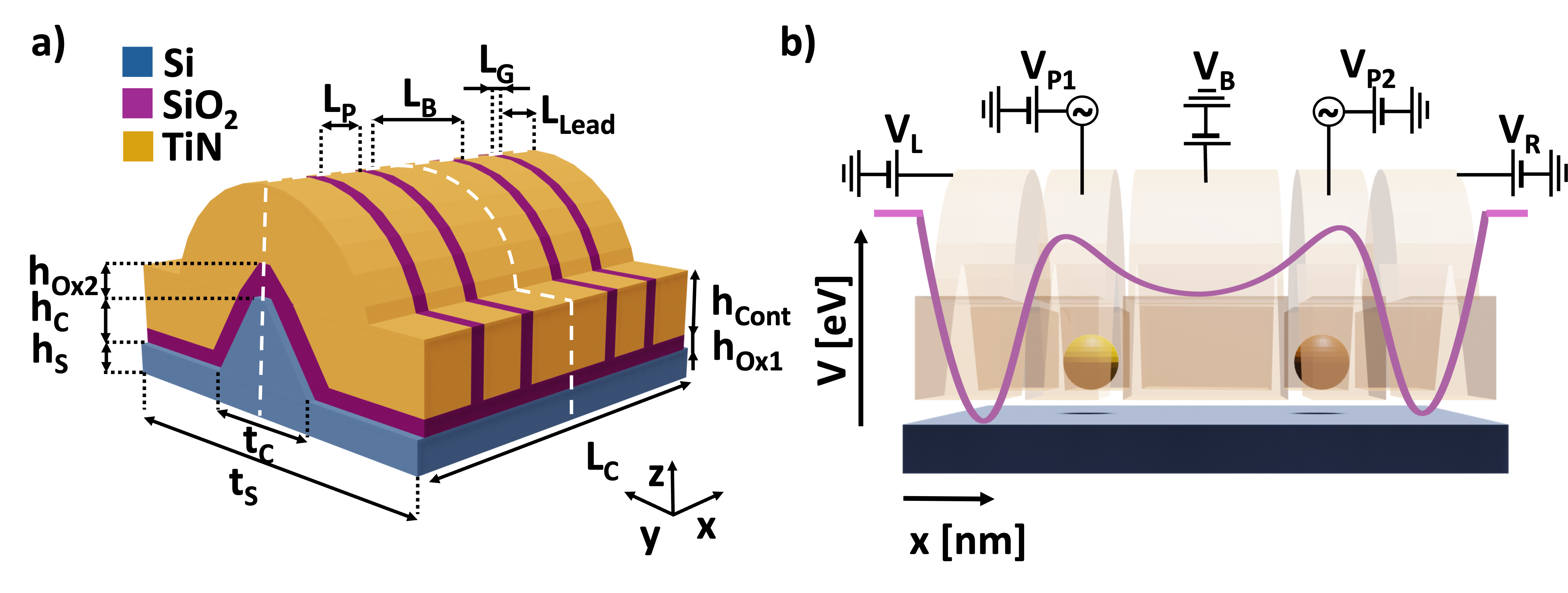}
\caption{\label{fig:device}(a) Illustration of the quantum device simulated in this work. It is a Si FinFET with a triangular channel surrounded by a SiO$_2$ oxide layer and operated with five TiN metallic gates that are shown in yellow. The SiO$_{2}$ layer is colored in purple and the Si regions in blue. The structure has the following dimensions: $L_{c}=95$~nm, $L_{Lead}=14$~nm, $L_{P}=12$~nm, $L_{G}=4$~nm, $L_{B}=27$~nm, $h_{S}=8$~nm, $h_{Ox1}=4$~nm, $h_{Ox2}=8$~nm, $h_{Cont}=20$~nm, $h_{C}=21$~nm, $t_{C} = 34$~nm, $t_{c2}=28$~nm, and $t_{S}=72$~nm. (b) Side view of the device in (a) showing the Si channel (transparent), the gate-induced QDs (yellow and brown spheres), and the substrate (opaque blue). The purple curve represents the average valence band variation throughout the transistor channel. It is controlled by three metallic gates (orange): the plunger gates $V_{P1/P2}$ to which an AC signal $V_{AC}$ can be applied and the left and right lead gates with their bias of $V_L$ and $V_R$, respectively.}
\end{figure}

\section{\label{sec:Methodology}Methodology}

\begin{figure*}[t]
\includegraphics[width=\textwidth]{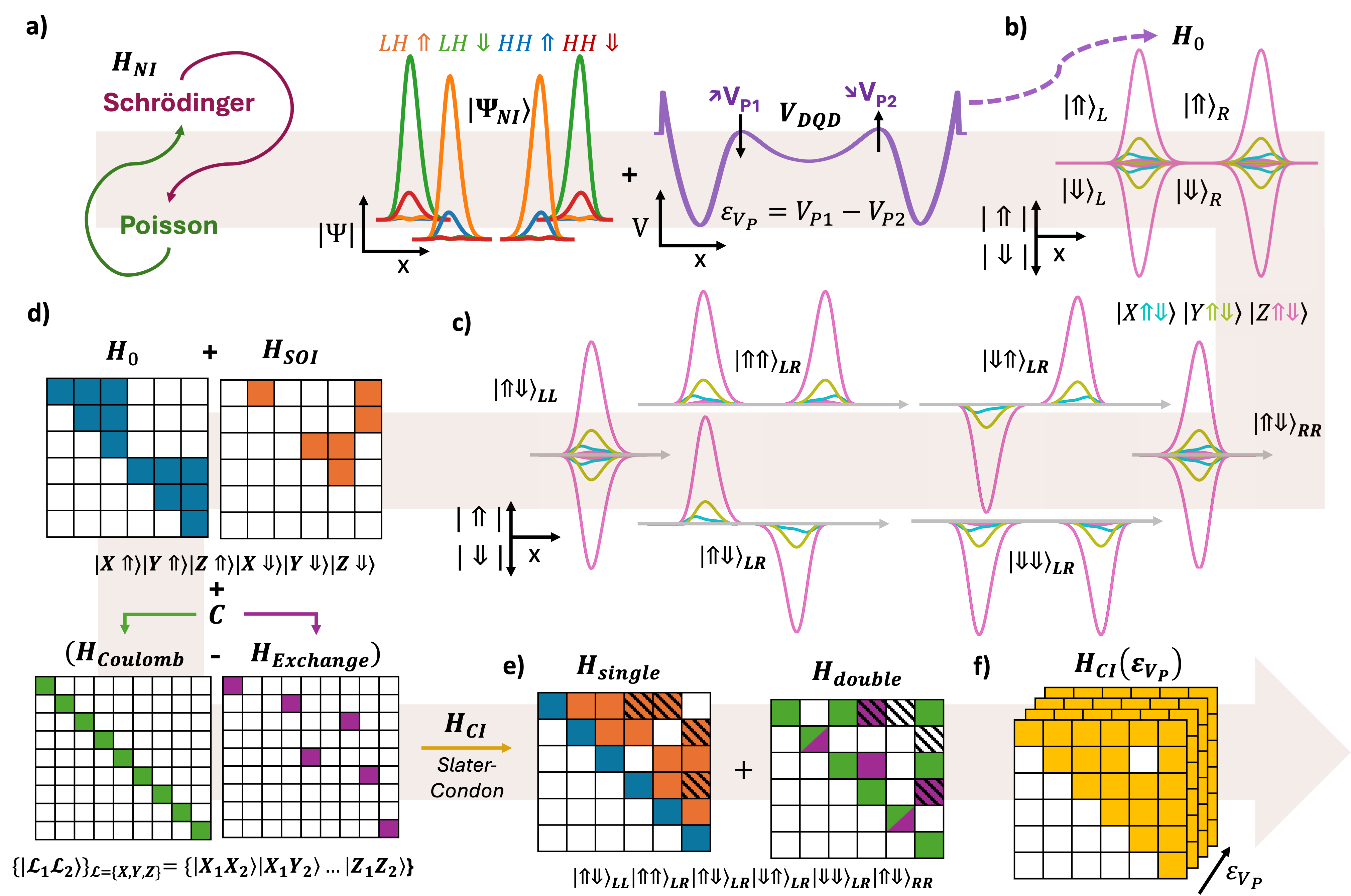}
\caption{\label{fig:methodology}Creation of a two-particle Hamiltonian through the CI method and a minimal single-particle basis set at zero magnetic field. (a) Self-consistent loop between the Schr\"odinger and Poisson equations. The first two doubly-degenerate hole eigenstates $|\Psi_{NI}^0\rangle$ and $|\Psi_{NI}^1\rangle$ with energy above the Fermi level obtained from diagonalizing Eq.~(\ref{eq:HNI}) are represented in terms of their band-mixing ($SO$ contribution omitted for visibility), while the converged electrostatic potential $V_{DQD}$ is shown as purple line. (b) First two doubly-degenerate localized eigenstates ($|\Uparrow\rangle_L/|\Downarrow\rangle_L$ and $|\Uparrow\rangle_R/|\Downarrow\rangle_R$) of Eq.~(\ref{eq:H0}) in absence of spin-orbit interaction, as obtained with our self-consistent SP solver. (c) Construction of the minimal basis set in terms of Slater determinants collected from all possible combinations of the single-particle states from (b). A total of 6 states labeled $|\Uparrow\Downarrow\rangle_{LL}$, $|\Uparrow\Downarrow\rangle_{LR}$, $|\Uparrow\Uparrow\rangle_{LR}$, $|\Downarrow\Downarrow\rangle_{LR}$, $|\Downarrow\Uparrow\rangle_{LR}$, and $|\Uparrow\Downarrow\rangle_{RR}$ are created. They constitute the basis expansion of the two-particle wavefunction from Eq.~(\ref{eq:Phi}). (d) Illustration of the non-zero elements of the Hamiltonian operators entering Eq.~(\ref{eq:H_DQDLS}) with their respective basis. Due to the hermicity of all the operators, only the upper triangular part of their matrix representations are displayed. $H_0$ and $H_{SOI}$ from Eq.~(\ref{eq:H0}) and Eq.~(\ref{eq:H1}), respectively, share the same $|\mathcal{L}S\rangle$ basis as $H_{DKK}$ from Eq.~(\ref{eq:LK}). On the other hand, the basis of $H_{Coulomb}$ and $H_{Exchange}$ given in Eq.~(\ref{eq:Cexplicit}) is the set of all two-orbital combinations arising from the product of the two single-particle states constituting the Slater determinant. (e) Same as (d) after application of the Slater-Condon rules to express Eq.~(\ref{eq:H_DQDLS}) in the Slater determinants basis of (c). The single-particle Hamiltonians $H_0$ and $H_{SOI}$ are grouped into $H_{single}=H_0+H_{SOI}$, the two-particle Hamiltonians $H_{Coulomb}$ and $H_{Exchange}$ into $H_{double}=H_{Coulomb}-H_{Exchange}$. The dashed entries correspond to Slater determinant pairs that share one particle state but at different position. According to the Slater-Condon rules \cite{slater1929theory,szabo2012modern,condon1930theory}, these entries are multiplied by -1 in the final configuration interaction $H_{CI}$ Hamiltonian. (f) Repetition of the steps highlighted in sub-plots (b), (c), and (d) for different $V_{DQD}$ to construct the corresponding $H_{CI}(\varepsilon_{V_{P}})$ at different detuning values $\varepsilon_{V_{P}}$. The diagonalization of each of these matrices yields the DQD energy spectrum $E(\varepsilon_{V_{P}})$.}
\end{figure*}

\subsection{Device Structure}

The quantum device under investigation in this study is presented in Fig.~\ref{fig:device}. It draws inspiration from the experimental FinFET structure previously reported in Ref.~\cite{geyer2024anisotropic}. A triangular Si channel with a height of 28~nm and a width of 15~nm is placed on a lightly $p$-doped (10$^{14}$ cm$^{-3}$), 8~nm thick Si substrate. At a cryogenic temperature of 1.5 K, two quantum dots can be electrostatically formed through the conjoint action of the TiN metallic gates. These gates are placed on top of an 8~nm thick SiO$_{2}$ insulating layer and equally spaced with a 4~nm thick layer of the same material. At both device extremities, two 14~nm long side gates ($V_{L,R}$) define extended reservoirs with high hole concentrations. Finally, two plunger gates ($V_{P1}$ and $V_{P2}$) of length 10~nm each determine the QD locations. A DC bias difference can be applied between $V_{P1}$ and $V_{P2}$ to induce an energy difference called detuning ($\varepsilon_{V_P}$) between the hole energy states of both dots, which host the qubits. On top of that, AC voltages can be used to manipulate the qubits. The central gate controls the overlap between the wavefunctions of the right and left qubit, which, together with the detuning, sets the magnitude of the exchange coupling between them. The device axes ($x$: channel axis, $y$ and $z$: directions of confinement) are oriented along the principal crystallographic directions, i.e., $x=[110]$, $y=[\bar110]$, and $z=[001]$. As compared to the experimental FinFET of \cite{geyer2024anisotropic}, the length of the source and drain extensions was reduced from 40~nm down to 14~nm to minimize the computational burden. This scaling is not expected to influence the DQD system which forms at sufficiently long distance away from these reservoirs. The gaps between the gates (4~nm) as well as the cross-section dimensions remain the same as in experiments. Finally, to numerically stabilize the electrostatic potential in the source and drain extensions, the device was extended on both sides with two 2-nm-thick and highly $p$-doped (7e19~cm$^{-3}$) Si regions that ensure flat-band conditions.

\subsection{Schr\"odinger-Poisson Solver}

In this section, we introduce the simulation framework we developed to characterize the electro-magnetic properties of the hole spin qubits hosted by the DQD system formed inside the FinFET structure of Fig.~\ref{fig:device}. The Hamiltonian matrix in the Schr\"odinger equation is expressed in the $6\times 6$ k$\cdot$p method and discretized on a homogeneous finite-difference (FD) grid with spacing $dx$=$dy$=$dz$=0.5~nm for a total number of $\sim3.25\times 10^5$ discretization points \cite{luisier_full-band_2008}. It is assumed that the hole wavefunctions do not penetrate into the surrounding oxide layers. Dirichlet boundary conditions are applied everywhere on the surface of the Si channel, thus forcing the wavefunction to vanish there. These boundary conditions turn the Schr\"odinger equation into an eigenvalue problem. The resulting eigenenergies and eigenstates are utilized to compute the system's hole density, which is then passed to a three-dimensional FD Poisson solver. In the latter, the electrostatic potential is fixed at the gate location and can vary everywhere else (von Neumann boundary conditions). Poisson's equation is solved in the full 3D domain, which encompasses $\sim$1.5$\times$10$^6$ points. The calculated electrostatic potential is then fed back into the Schr\"odinger solver and the procedure is repeated till convergence is reached.

In terms of numerics, both our Schr\"odinger and Poisson kernels have been optimized to allow for the treatment of large simulation domains and to reduce the computational time. All eigenpairs of the Schrödinger equation are determined via the Rayleigh-Chebyshev subspace iteration method \cite{anderson2010rayleigh} that has been ported to both CPUs and GPUs. As compared to the standard ARPACK routines on CPUs \cite{lehoucq1998arpack}, a speed up by a factor of $\times$2.9 (CPUs) and $\times$20 (GPUs) is obtained with our implementation, making the investigation of the DQD system from Fig.~\ref{fig:device} a computationally affordable task.

Since spin qubit devices are typically operated at cryogenic temperatures, thermal deformation effects upon cooling must be accounted for as they have a strong impact on the DQD eigenenergies and eigenstates. In Ref.~\cite{bouquet2025simulation}, it was found that a contraction of the Si channel is the most likely scenario to occur when the FinFET is cooled down to 1.5 K. Therefore, the same conditions are applied here by fixing the bottom edge of the Si wafer, leading to an overall contraction of the material stack towards the substrate. The resulting non-homogeneous strain tensor is included into the $6\times 6$ k$\cdot$p Hamiltonian through the Pikus-Bir approach \cite{bir1974symmetry}. 

\subsection{\label{Exchange_coupling}Exchange Coupling}
A pair of interacting spins can coexist, either forming a singlet state with spin angular momentum $S=0$ or a triplet state with angular momentum $S=1$. Since holes are fermions, they must obey Pauli's exclusion principle, which dictates that the total wavefunction of a given system must be antisymmetric. If this wavefunction is decomposed into a product of spatial and spin states, singlets (triplets) have an antisymmetric (symmetric) spin part, whereas their spatial part must be symmetric (antisymmetric) to fulfill the aforementioned antisymmetry criterion. Because of Coulomb repulsion, singlet (the holes involved are close to each other) and triplet (holes are located further apart) states have different energies that are separated by the so-called exchange coupling $J$:
\begin{eqnarray}
J=E_{Triplet}-E_{Singlet}.
\label{eq:J}
\end{eqnarray}

Computing the exchange coupling energy $J$ between two interacting quantum dots consists of diagonalizing the two-particle Hamiltonian given by
\begin{subequations}
\label{eq:HDQD}
\begin{eqnarray}
H_{DQD}=H_{NI}+C,
\end{eqnarray}
\begin{eqnarray}
\label{eq:HNI}
H_{NI}=H_{LK}+H_{PB}+V_{DQD},
\end{eqnarray}
\end{subequations}
where the non-interacting part $H_{NI}$ is made, in our case, of the well-known $6\times6$ Luttinger-Kohn $k\cdot p$ Hamiltonian $H_{LK}$\cite{luttinger1956quantum}, the Pikus-Bir Hamiltonian $H_{PB}$ \cite{bir1974symmetry}, and the gate-induced electrostatic potential hosting both QDs $V_{DQD}$. The Coulomb interaction $C$ between two dots positioned at $\boldsymbol{r_1}$ and $\boldsymbol{r_2}$ is defined as
\begin{eqnarray}
C=\frac{e}{\varepsilon|\boldsymbol{r_1}-\boldsymbol{r_2}|}.
\label{eq:Coulomb}
\end{eqnarray}
It depends on the total permittivity $\varepsilon=\varepsilon_0\varepsilon_{Si}$, which is the product of the vacuum ($\varepsilon_0$) and relative permittivity of Si ($\varepsilon_{Si}$). In Eq.~(\ref{eq:HDQD}), the Luttinger-Kohn Hamiltonian is generally expressed in the basis of the total angular momentum $\mathscr{J}$ and its projection onto the $z$ axis, $m_j$, i.e., $|\mathscr{J},m_j\rangle$=$|\frac{3}{2},+\frac{3}{2}\rangle$, $|\frac{3}{2},+\frac{1}{2}\rangle$, $|\frac{3}{2},-\frac{1}{2}\rangle$, $|\frac{3}{2},-\frac{3}{2}\rangle$, $|\frac{1}{2},+\frac{1}{2}\rangle$, and $|\frac{1}{2},-\frac{1}{2}\rangle$. Through a rotation $U$, $H_{LK}$ can be decomposed into two parts \cite{willatzen2009kp}:
\begin{eqnarray}
UH_{LK}U^{\dagger}=H_{DKK}+H_{SOI}.
\label{eq:LK}
\end{eqnarray}
In Eq.~(\ref{eq:LK}), $H_{DKK}$ is the Dresselhaus-Kittel-Kip Hamiltonian. It is expressed in the basis of the orbital angular momentum $\mathcal{L}$ and the spin angular momentum $S$, i.e., $|\mathcal{L}S\rangle$=$|X\Uparrow\rangle$, $|Y\Uparrow\rangle$, $|Z\Uparrow\rangle$, $|X\Downarrow\rangle$, $|Y\Downarrow\rangle$, and $|Z\Downarrow\rangle$. The term labeled $H_{SOI}$ contains the spin-orbit coupling. Note that $H_{PB}$ in Eq.~(\ref{eq:HNI}) must also be rotated according to the matrix $U$, thus giving rise to a new term $H'_{PB}=UH_{PB}U^{\dagger}$. The DQD Hamiltonian $H_{DQD}$ in Eq.~\ref{eq:HDQD} can then be rewritten as
\begin{eqnarray}
\label{eq:H_DQDLS}
H_{DQD}=H_0+H_1,
\end{eqnarray}
with
\begin{subequations}
\label{eq:H0}
\begin{eqnarray}
\label{eq:realH0}
H_0=H_{DKK}+V_{DQD}+H'_{PB},
\end{eqnarray}
\begin{eqnarray}
\label{eq:H1}
H_1=H_{SOI}+C.
\end{eqnarray}
\end{subequations}
Importantly, $V_{DQD}$ and $C$ remain unchanged by the action of $U$. In Eq.~(\ref{eq:H_DQDLS}), $H_0$ contains only single-particle terms, contrary to $H_1$, which mixes single- and double-particle contributions. 

In the CI framework, the Hamiltonian in Eq.~(\ref{eq:H_DQDLS}) is usually expanded in terms of Slater determinants which, in our case, represent two-particle wavefunctions $|\Phi^k_{DQD}\rangle$ constructed from an orthonormal basis set of single-particle wavefunctions $|\varphi_i\rangle$ \cite{slater1929theory}: 
\begin{eqnarray}
\label{eq:Phi}
|\Phi_{DQD}^k\rangle=\sum_{i<j}^{K}{c_{ij}^k|\varphi_i\varphi_j\rangle},
\end{eqnarray}
where $\varphi_i$ is the $i^{th}$ eigenstate with energy $E_i$ of the non-interacting system described by Eq.~(\ref{eq:realH0}), $c_{ij}^k$ represents the expansion coefficient of the $k^{th}$ state of the coupled system, and $K$ is the size of the basis expansion. By combining Eqs.~(\ref{eq:LK}) to (\ref{eq:H1}), the original system in Eq.~(\ref{eq:HDQD}) can be practically solved, which returns the $K$ vectors of coefficients $c_{ij}^k$ and their corresponding energies as $\{c_{ij}^k,E_{DQD}^k\}_{k=0}^{k=K-1}$. Finally, the two-particle wavefunction $|\Phi_{DQD}^k\rangle$ can be reconstructed from the coefficient expansions and the exchange coupling $J$ in Eq.~(\ref{eq:J}) can be extracted.

Concretely, we follow the methodology steps outlined in Fig.~\ref{fig:methodology}. We first perform a self-consistent SP simulation with the non-interacting Hamiltonian matrix $H_{NI}$ in Eq.~(\ref{eq:HNI}), assuming that two holes are present, those corresponding to the bonding and anti-bonding states. As a result, we obtain the electrostatic potential $V_{DQD}$ of the DQD structure as well as $N$ eigenpairs $\{|\Psi_{NI}^{n}\rangle,E_{NI}^{n}\}_{n=0}^{n=N-1}$. 
To ensure the presence of two holes, only $E_{NI}^{0}$ and $E_{NI}^{1}$ lie above the Fermi level so that $|\Psi_{NI}^{0}\rangle$ and $|\Psi_{NI}^{1}\rangle$ are unoccupied, doubly-degenerate states (Fig.~\ref{fig:methodology}(a)). The same approach is repeated for a broad range of detuning $\varepsilon_{V_P}$ between the plunger gates. Each simulation returns a different electrostatic potential $V_{DQD}(\varepsilon_{V_P})$. 

Next, to create an orthonormal basis of single-particle states, we diagonalize $H_0$ in Eq.~(\ref{eq:H0}) with the pre-computed electrostatic potential $V_{DQD}$. Because spin-orbit coupling is not included in $H_0$, $H_{DKK}$ and $H'_{PB}$ are made of two identical 3$\times$3 blocks $H_{DKK,3\times3}$ and $H'_{PB,3\times3}$ that, when diagonalized, lead to the same eigenpairs. It is therefore sufficient to retain only one of these blocks. Such a spin-independent approach is more convenient to describe systems with unpaired electrons, e.g., triplet states, as it automatically produces restricted Hartree-Fock (RHF) determinants where the orbital part of the doubly-degenerate eigenstates is exactly the same for the ``spin-up'' and ``spin-down'' states, thus avoiding spin contamination \cite{szabo2012modern}. As an example, if we consider only the first two doubly-degenerate eigenstates of $H_0$, we obtain the following minimal single-particle basis $\{|\varphi_{i}\rangle\}_{i=0}^{i=3}$ which we have relabeled $|\Uparrow\rangle_L$, $|\Downarrow\rangle_L$, $|\Uparrow\rangle_R$, and $|\Downarrow\rangle_R$ in Fig.~\ref{fig:methodology}(b). Here, $\Uparrow$/$\Downarrow$ refer to the spin state, $L/R$ to the position of the hole (left or right) QD. Note that the so far left-out spin-orbit coupling is included in $H_1$ in Eq.~(\ref{eq:H1}). It will be reintroduced later in the procedure.
 
As next step, a collection of Slater determinants is constructed from the $\{|\varphi_{i}\rangle\}_{i=0}^{i=M-1}$ eigenstates. The number of possible combinations of these states, $K$, is given by the binomial coefficient:
\begin{eqnarray}
\label{eq:K}
K=\binom{M}{n_p},
\end{eqnarray}
$M$ being the number of single-particle eigenstates included and $n_p$ the number of particles in the system, two in the present DQD configuration. For example, if $M=4$, there exist $K=6$ different Slater determinants $|\varphi_{i}\varphi_j\rangle$=$|\Uparrow\Downarrow\rangle_{LL}$, $|\Uparrow\Downarrow\rangle_{LR}$, $|\Uparrow\Uparrow\rangle_{LR}$, $|\Downarrow\Downarrow\rangle_{LR}$, $|\Downarrow\Uparrow\rangle_{LR}$, and $|\Uparrow\Downarrow\rangle_{RR}$, as depicted in Fig.~\ref{fig:methodology}(c). This basis, together with the $c_{ij}^k$ coefficients from Eq.~(\ref{eq:Phi}), is used to expand the Hamiltonian $H_{DQD}$ in Eq.~(\ref{eq:H_DQDLS}), bringing back spin-orbit coupling and Coulomb interactions through $H_1$. Through application of the Slater-Condon rules, all components of $H_0$, $H_{SOI}$, and $C$, can be straightforwardly calculated, which leads to the configuration-interaction Hamiltonian $H_{CI}$ \cite{szabo2012modern,slater1929theory,condon1930theory}. In Fig. \ref{fig:methodology}(d) we represent the different contributions, separating those depending on single- and double-particle basis elements. Finally, $H_{CI}=H_{single}+H_{double}$ is built in Fig.~\ref{fig:methodology}(e). Focusing first on the single-particle Hamiltonian components, $H_0$ and $H_{SOI}$, the bra-ket of two Slater determinants is only non-zero if they possess at least one single-particle state in common. In this case:
\begin{eqnarray}
\label{eq:single}
\langle\varphi_i\varphi_j|h|\varphi_i\varphi_l\rangle=\langle\varphi_j|h|\varphi_l\rangle,
\end{eqnarray}
and for identical determinants:
\begin{eqnarray}
\label{eq:identical}
\langle\varphi_i\varphi_j|h|\varphi_i\varphi_j\rangle=\langle\varphi_i|h|\varphi_i\rangle+\langle\varphi_j|h|\varphi_j\rangle,
\end{eqnarray}
where $h$ represents any single-particle Hamiltonian. Since the single-particle states are the eigenvectors of $H_0$, Eq.~(\ref{eq:single}) is zero if $h=H_0$, leading to only the on-site (blue) contributions to $H_{CI}$ in Fig.~\ref{fig:methodology}(e). When $h=H_{SOI}$, the Slater-Condon rules produce the non-diagonal orange contributions to $H_{single}$, owing to the fact that the spin-orbit interaction between two identical wavefunctions is zero.

Turning now to the two-particle contributions, $H_{double}$, the expansion of the Coulomb interaction $C$ can be divided into two parts denoted $H_{Coulomb}$ ($H_{Cou.}$) and $H_{Exchange}$ ($H_{Exch.}$). In our two-hole system, they are defined as:
\begin{equation}
\label{eq:CEx}
\begin{split}
\langle\varphi_i\varphi_j|C|\varphi_k\varphi_l\rangle = & \langle\varphi_i\varphi_j|H_{Cou.}|\varphi_k\varphi_l\rangle \\
& - \langle\varphi_i\varphi_j|H_{Exch.}|\varphi_l\varphi_k\rangle,
\end{split}
\end{equation}
regardless of the number of shared single-particle states. Nevertheless, the Coulomb interaction only acts on the spatial part of the wavefunction and leaves the spin component unaffected. As a consequence, if the spin orientation between two interacting single-particle states is anti-parallel, then the corresponding bra-ket term in Eq.~(\ref{eq:CEx}) is zero. Therefore, selection rules based on the spin asymmetry between Slater determinants can be deduced so that only non-zero elements are computed: 
\begin{subequations}
\label{eq:spinHc}
\begin{eqnarray}
\langle\varphi_i\varphi_j|H_{Cou.}|\varphi_k\varphi_l\rangle\neq 0 \text{ only if} \\\sigma(\varphi_i)=\sigma(\varphi_k) \text{ and }\sigma(\varphi_j)=\sigma(\varphi_l),
\end{eqnarray}
\begin{eqnarray}
\label{eq:spinHe}
\langle\varphi_i\varphi_j|H_{Exch.}|\varphi_l\varphi_k\rangle\neq 0 \text{ only if} \\\sigma(\varphi_i)=\sigma(\varphi_l) \text{ and }\sigma(\varphi_j)=\sigma(\varphi_k),
\end{eqnarray}
\end{subequations}
with $\sigma(\varphi_i)=\{\Uparrow,\Downarrow\}$.
These properties are exemplified in Fig.~\ref{fig:methodology}(e) where the green and purple entries mark the non-zero elements of $H_{Coulomb}$ and $H_{Exchange}$, respectively. Furthermore, since the single-hole eigenstates $\{\varphi_{i}\}_{i=0}^{i=M-1}$ are expressed as linear combination of the $p$-orbitals $\{|X\rangle,|Y\rangle,|Z\rangle\}$, the product between the orbitals of the two single-particle states of each Slater determinant must be carried out. For instance, renaming $|\varphi_{0}\rangle$ as $|\Uparrow\rangle_L$ and $|\varphi_{1}\rangle$ as $|\Downarrow\rangle_L$, we create the Slater determinant $|\Uparrow\Downarrow\rangle_{LL}$. As depicted in Fig.~\ref{fig:methodology}(c) the determinant $|\Uparrow\Downarrow\rangle_{LL}$ is the product between $|\Uparrow\rangle_L=\sum_{|\mathcal{L}\rangle}c_{\Uparrow_L}^{|\mathcal{L}\rangle}|\mathcal{L}\Uparrow\rangle$ and $|\Downarrow\rangle_L=\sum_{|\mathcal{L}\rangle}c_{\Downarrow_L}^{|\mathcal{L}\rangle}|\mathcal{L}\Downarrow\rangle$, with $\mathcal{L}=\{|X\rangle,|Y\rangle,|Z\rangle\}$. Expanding the latter product results in an explicit sum over the following nine pair combinations: $|X\Uparrow X\Downarrow\rangle$, $|X\Uparrow Y\Downarrow\rangle$, $|X\Uparrow Z\Downarrow\rangle$, $|Y\Uparrow X\Downarrow\rangle$, $|Y\Uparrow Y\Downarrow\rangle$, $|Y\Uparrow Z\Downarrow\rangle$, $|Z\Uparrow X\Downarrow\rangle$, $|Z\Uparrow Y\Downarrow\rangle$, and $|Z\Uparrow Z\Downarrow\rangle$ with the corresponding coefficients $c_{\Uparrow_L}^{|\mathcal{L}\rangle*}c_{\Downarrow_L}^{|\mathcal{L}\rangle}$. The bra-ket combination between two Slater determinants thus implies the computation of $9\times9=81$ terms before their final summation to generate one unique non-zero entry. However, besides being spin independent, the Coulomb interaction is also intraband. That is, the operator $C$ cannot couple orbitals of different nature, e.g., $\langle X|$ with $|Y\rangle$, hence from the previous 81 terms only 9 remain, namely the ones with pair of determinants exhibiting the same orbitals (as well as their ordering) in the ``bra'' and the ``ket''. This additional selection rule is illustrated in Fig.~\ref{fig:methodology}(d) by the non-zero green and purple entries of $H_{Coulomb}$ and $H_{Exchange}$, respectively. Having now tagged all non-zero contributions to the two-particle Hamiltonian, we can explicitly rewrite the first term of Eq.~(\ref{eq:CEx}) as: 
\begin{eqnarray}
\label{eq:Cexplicit}
&\langle\varphi_i^\sigma\varphi_j^{\sigma'}|H_{Cou.}|\varphi_k^\sigma\varphi_l^{\sigma'}\rangle = \nonumber\\ 
&\frac{e}{\varepsilon} \iint \frac{\varphi^{\sigma*}_i(\boldsymbol{r_1})\varphi^{\sigma'*}_j(\boldsymbol{r_2})\varphi_k^\sigma(\boldsymbol{r_1})\varphi_l^{\sigma'}(\boldsymbol{r_2})}{|\boldsymbol{r_1}-\boldsymbol{r_2}|}
d\boldsymbol{r_1}\,d\boldsymbol{r_2}.
\end{eqnarray}
To compute $\langle\varphi_i^\sigma\varphi_j^{\sigma'}|H_{Cou.}|\varphi_k^\sigma\varphi_l^{\sigma'}\rangle$, we first introduce the potential energy:
\begin{eqnarray}
V^{\sigma'}_{jl}(\boldsymbol{r_2})=\frac{e}{\varepsilon}\int\frac{\varphi^{\sigma*}_j(\boldsymbol{r_2})\varphi_l^\sigma(\boldsymbol{r_2})}{|\boldsymbol{r_1}-\boldsymbol{r_2}|}\,d\boldsymbol{r_2}
\end{eqnarray}
and the complex density $\rho^{\sigma'}_{jl}=\varphi^{\sigma'*}_j(\boldsymbol{r_2})\varphi_l^{\sigma'}(\boldsymbol{r_2})$, where $\sigma$ represents the spin degree of freedom. Hence, $V_{jl}(\boldsymbol{r_2})$ can be determined by solving Poisson's equation:
\begin{eqnarray}
\nabla^2V^{\sigma'}_{jl}=-\frac{e}{\varepsilon}\rho^{\sigma'}_{jl}.
\end{eqnarray}
Altogether, this reduces Eq.~(\ref{eq:Cexplicit}) to
\begin{eqnarray}
\label{eq:Cfinal}
\langle\varphi_i^\sigma\varphi^{\sigma'}_j|H_{Cou.}|\varphi_k^\sigma\varphi^{\sigma'}_l\rangle=\int~V^{\sigma'}_{jl}(\boldsymbol{r_1})\rho^\sigma_{ik}(\boldsymbol{r_1})\,d\boldsymbol{r_1}.
\end{eqnarray}
The exact same procedure from Eqs.~(\ref{eq:Cexplicit}) to (\ref{eq:Cfinal}) can be followed to determine the exchange term, $H_{Exch.}$, in Eq.~(\ref{eq:CEx}).
Finally, with the previously computed electrostatic potentials at different detuning, $V_{DQD}(\varepsilon_{V_P})$, steps (b) to (e) in Fig.~\ref{fig:methodology} are repeated to generate a collection of detuning-dependent CI Hamiltonians, $H_{CI}(\varepsilon_{V_P})$ (see Fig.~\ref{fig:methodology}(f)).

\section{\label{sec:Results}Results \& Discussion}

\begin{figure}[t]
\includegraphics[width=\columnwidth]{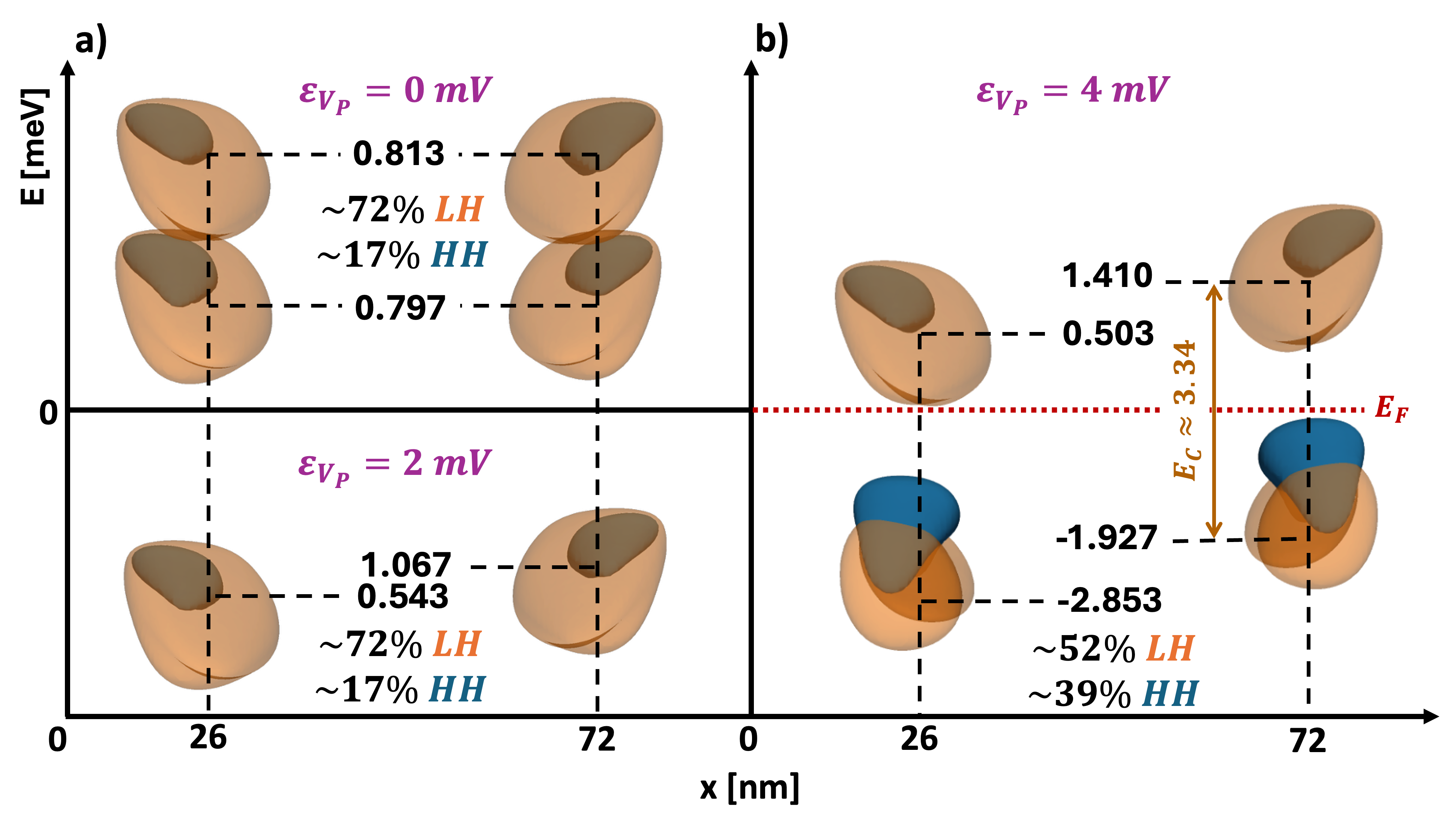}
\caption{\label{fig:bandmixing} Isosurface of the heavy-hole (blue) and light-hole (orange) contributions to the hole density in the device from Fig.~\ref{fig:device} at $V_{P1/P2}=484$~mV, $V_B=690$~mV and $V_{L/R}=950$~mV. The $x$-coordinates of the left and right dots are given on the horizontal axis. (a) Bonding ($E_{\varepsilon_{V_P}=0\textrm{~mV}}$=0.813 meV and $E_{\varepsilon_{VP}=2\textrm{~mV}}$=1.067 meV) and anti-bonding state ($E_{\varepsilon_{V_P}=0\textrm{~mV}}$=0.797 meV and $E_{\varepsilon_{V_P}=2\textrm{~mV}}$=0.543 meV) in case of a symmetric DQD system (top) and with a finite detuning $\varepsilon_{V_P}$=1~mV between both QDs (bottom). (b)  Same as in (a), but for $\varepsilon_{V_P}$=1.5~mV, including the first ($E_{\varepsilon=4\textrm{~mV}}$=-1.927 meV) and second ($E_{\varepsilon=4\textrm{~mV}}$=-2.853 meV) excited states, which are situated below the Fermi level $E_F$=0 eV. The omitted band-mixing for the bonding and anti-bonding state is the same as in (a). Besides, the brown double-arrow between the bonding and first excited state depicts the charging energy $E_C$=3.34 meV. For visibility purposes, the energy separation between the states is scaled differently in sub-plot (a) and (b).}
\end{figure}

\begin{figure*}
\includegraphics[width=\textwidth]{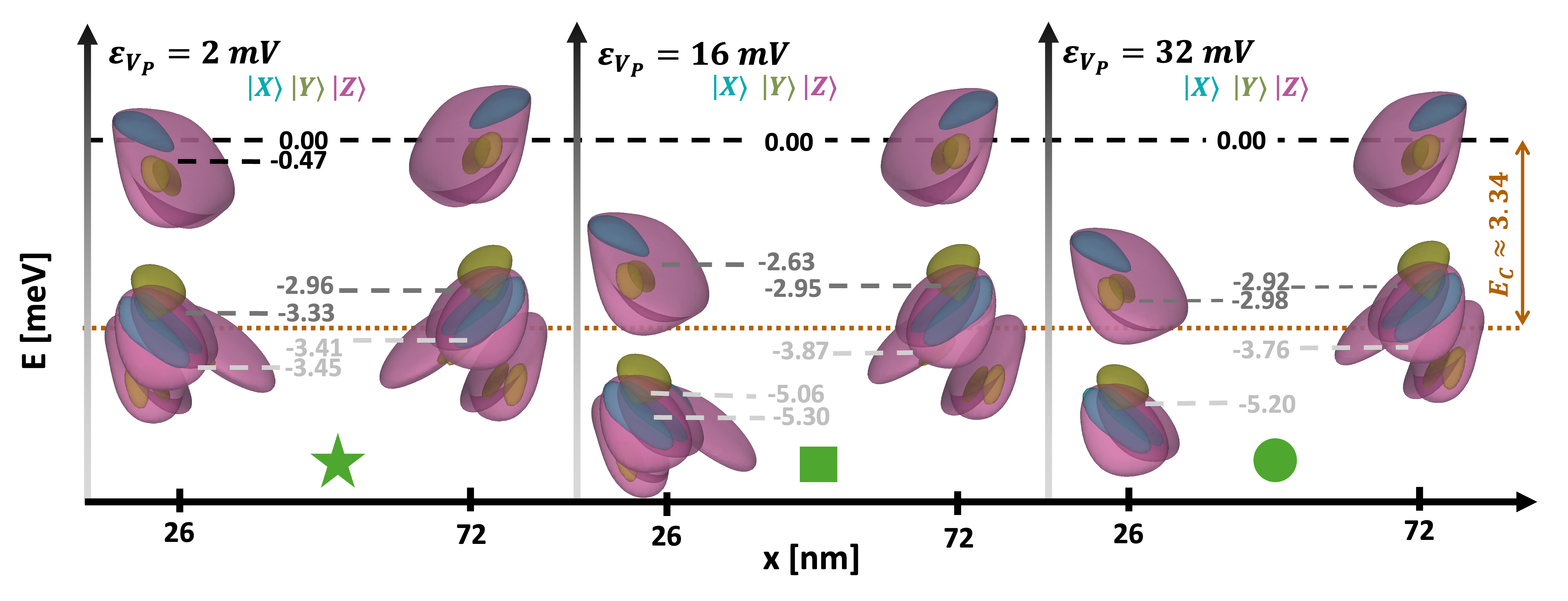}
\caption{\label{fig:bandmixingDKK} Isosurface of the doubly degenerate $X$ (cyan), $Y$ (green), and $Z$ (pink) components of the first six highest single-particle state, weighted by their relative probability, as obtained by solving Eq.~(\ref{eq:H0}) for $V_B=690$~mV, $V_{L/R}=0.95$~mV, and the following voltage pairs:$V_{P1}/V_{P2}=485/483 Bowen$~mV $\iff\varepsilon_{V_P}=2$~mV (star), $V_{P1}/V_{P2}=492/476$~mV $\iff\varepsilon_{V_P}=16$~mV (triangle), and $V_{P1}/V_{P2}=500/468$~mV $\iff\varepsilon_{V_P}=32$~mV (circle, sixth doubly-degenerate single-particle state not shown). The $x$-coordinates of the left and right dots are given on the horizontal axis. The energy of the ground state is fixed at 0 eV, whereas that of the excited states is shifted according to their energy separation to the ground state. The reported energy levels are color-coded from black for the high-energy ground state to light gray for the last low-energy excited state. Moreover, the charging energy $E_C$, as estimated in Fig.~\ref{fig:bandmixing}, is reported. It indicates the energy threshold above which the single-particle eigenstates are truncated from the CI space.}
\end{figure*}

\begin{figure*}
\includegraphics[width=\textwidth]{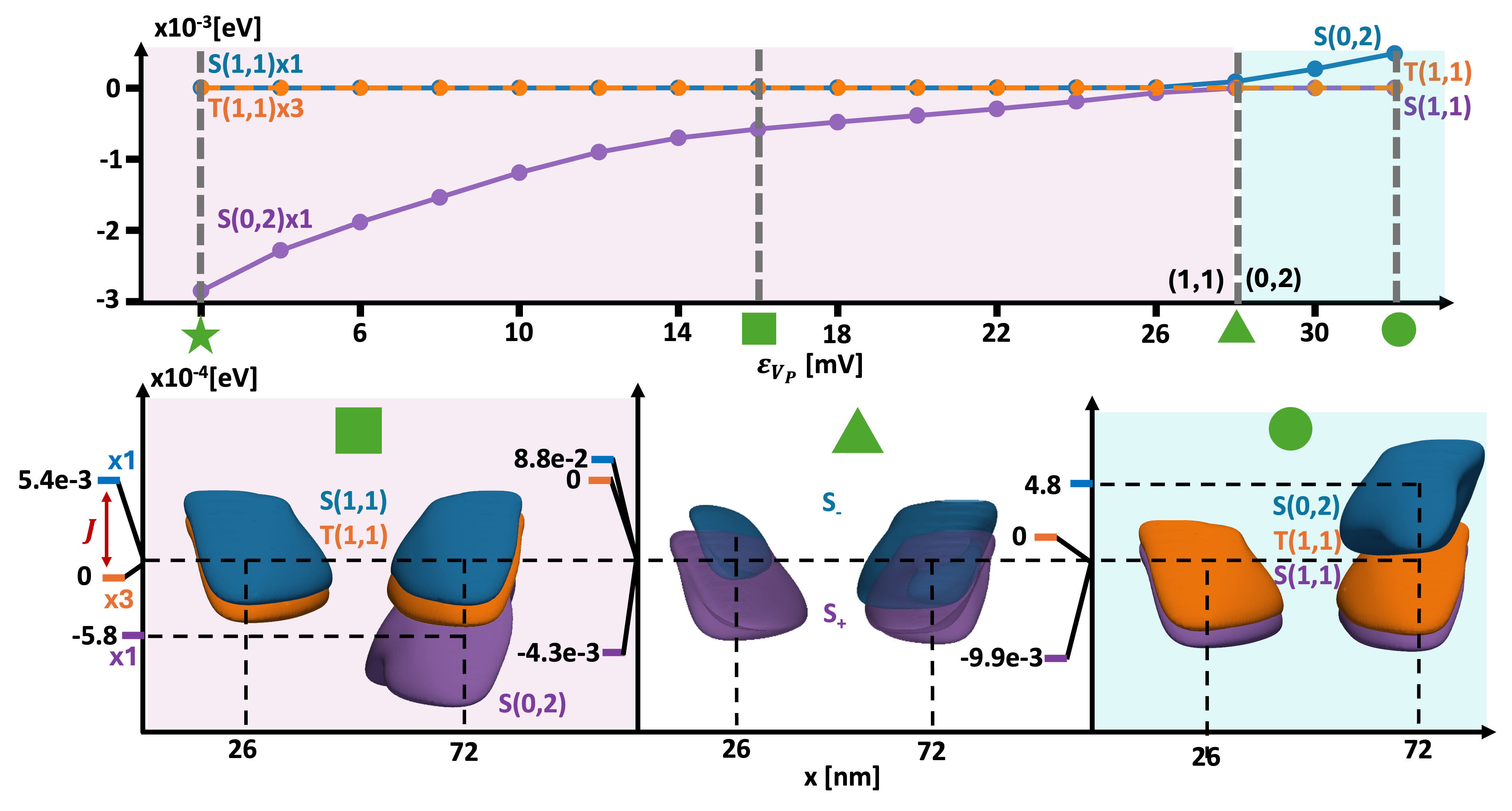}
\caption{\label{fig:iso_eigenvectors}(top) Energy spectrum of the 5-gate triangular FinFET from Fig.~\ref{fig:device} obtained at $V_B=690$~mV, $V_{L/R}=950$~mV, and $V_{P1}(V_{P2})$ ranging from 484 (484)~mV to 500 (468)~mV. The five highest double-particle states are represented as a function of the detuning $\varepsilon_{V_P}$=$V_{P1}$-$V_{P2}$. They correspond to $S(1,1)$ (highest energy singlet), $T(1,1)$ (three-fold degenerate triplet), and $S(0,2)$ (singlet). In the pink region labeled (1,1), the highest-energy configuration is the one with one hole per QD, while in the (0,2) blue region, the energetically most favorable configuration is that where two holes are located in the right QD. The green symbols refer to detuning situations that are further investigated in this paper. (bottom) Isosurface of the $S(1,1)$, $T(1,1)$ (triply-degenerate), and $S(0,2)$ states from the top pane after being reconstructed from the coefficients in Eq.~(\ref{eq:Phi}). The same color code is used to represent the five highest double-particle states at different detuning values, namely those marked by green symbols. The degeneracy of the two-particle eigenstates is indicated on the far left axis and remains the same for all $\varepsilon_{V_P}$ values. The exchange coupling $J$ is indicated in red as the energy separation between the blue singlet and orange triplet states. Compared to Eq.~(\ref{eq:J}), the exchange coupling value is multiplied by -1 owing to the hole nature of the system.}
\end{figure*}

\subsection{Double Quantum Dot Formation}
All simulations are performed at cryogenic temperature $T$=1.5 K and cooling-induced stress is always included. Under these conditions and with the material parameters from Appendix \ref{A:parameters}, a symmetric DQD system is created when $V_{P1}$=$V_{P2}$=0.484~V, $V_{B}$=0.690~V, and $V_{L}$=$V_{R}$=0.95~V. Figure~\ref{fig:bandmixing} shows the band-mixing between the dominant light-hole (LH, orange) and heavy-hole (HH, blue) populations in the form of isosurfaces. The LH and HH joint contributions encompass 98\% of the total DQD charge. At zero detuning ($\varepsilon_{V_P}$=0~mV), the single-particle states obtained from the self-consistent SP simulations are delocalized below both plunger gates. However, due to numerical noise around the bias symmetry point, the eigenfunctions of the DQD system are not exactly symmetric. This is indeed evidenced by the slightly asymmetrical distribution of the isosurfaces in Fig.~\ref{fig:bandmixing}(a), in particular by the blue HH density that shows clear uneven distribution between the left and right dot. In all cases, the light-hole contribution to the bonding and anti-bonding states remains largely superior to the heavy-hole one (82\% vs. 15\%), with about 3\% coming from the spin-orbit coupling. Such a distribution is indeed to be expected considering the presence of two strong directions of confinement and of cooling-induced compressive strain, both giving rise to a strong light-hole population \cite{csontos2009spin,moratis2021light,kloeffel2011strong}. 

We present in Fig.~\ref{fig:bandmixingDKK} the isosurface of the orthonormal basis set computed from the $3\times 3$ Hamiltonian of Eq.~(\ref{eq:H0}) as a function of its $|X\rangle$, $|Y\rangle$, and $|Z\rangle$ components as well as of the detuning $\varepsilon_{V_P}$. The energy levels of all single-hole states (six per detuning) are reported on a gray scale, from the two high-energy ground states in black down to the four low-energy excited states in gray.
Due to the bias difference between the plunger gates, the hole states are localize below either one of these regions, with the four higher-energy states displaying a narrower distribution of their charge density compared to the two lower ones pictured in light-gray. At low detuning ($0<\varepsilon_{V_P}<14 Bowen$~mV), the two ground states located at 0.0 meV and -0.47 meV are well separated from the next pair of states at -2.96 meV and -3.33 meV. Hence, in the most compact theoretical interaction model, a minimal basis formed only by the two highest-energy hole ground states, with spin up and spin down, is sufficient to describe our DQD system, leading exactly to the six determinants shown in Fig.~\ref{fig:methodology}(c). As the detuning increases, the energy spacing between the different levels changes so that for regions of moderate to strong detuning ($\varepsilon_{V_P}>$15~mV), the basis expansion should include the middle-energy states (gray) and be formed of four, doubly-degenerate states. However, the light-gray hole states with the lowest energy remain distant enough to be ignored for all detuning values. 

These observations suggest that the CI space can be truncated to retain only the primary contributors to the exchange interaction. As a physically meaningful energy threshold, we use the charging energy $E_C$. It is defined as the energy required to add an extra hole to the system and corresponds to the size of the diamonds in transport measurement stability diagrams also called ``honeycomb'' plots \cite{zwanenburg2013silicon}. Since the contribution of states with an energy superior to $E_C$ is limited, applying a cut-off to this quantity ensures that the CI model only includes states that are physically accessible during qubit manipulations. In our triangular FinFET, we estimate that $E_C$ is $\sim3.34$ meV, which corresponds to the energy difference between the bonding ground state and the first excited state below the Fermi level $E_F$, as depicted in Fig.~\ref{fig:bandmixing}(b). Note that the truncation of the states is realized by applying a Fermi-Dirac function as a filter at $E_F=E_C$ with a temperature $T<1.5$ K as input parameter. 

To illustrate the benefit of such a reduced basis, we plot in Fig.~\ref{fig:iso_eigenvectors} the energy spectrum of the DQD system from Fig.~\ref{fig:device} with respect to the detuning parameter $\varepsilon_{V_P}$, as obtained from the diagonalization of the CI Hamiltonian $H_{DQD}$ in Eq.~(\ref{eq:H_DQDLS}). By keeping only the eigenstates with energy above $E_C$, namely the four highest-energy single-hole states, $K=28$ determinants are built. By applying a detuning voltage between both dots, a transition from a symmetric configuration labeled (1,1), where each dot hosts one hole, to an asymmetric one called (0,2), if $\varepsilon_{V_P}>0$, or (2,0), if $\varepsilon_{V_P}<0$, where one of the dots is charged with two holes, can be observed. Note that the energy spectrum in Fig.~\ref{fig:iso_eigenvectors} is reversed compared to the electron case because of the hole nature of our system. Hence, the commonly singly- or doubly-occupied two-electron states become singly- or doubly-unoccupied two-hole states. 

Without any applied magnetic field, the theory postulates that the ground state (blue curve) of our DQD system is either a delocalized singlet state $S(1,1)$ ((1,1) pink area in Fig.~\ref{fig:iso_eigenvectors}) or a localized singlet state with a doubly-unoccupied dot $S(0,2)$ $(S(2,0))$ in the (0,2) ((2,0)) region (blue area) \cite{mattis2012theory}. The first excited state, at zero magnetic field and in both regions, is the triply-degenerate triplet state $T(1,1)$ (orange curve) whose energy separation with the $S(1,1)$ state inside the (1,1) region corresponds to the exchange coupling $J$ in Eq.~(\ref{eq:J}). This quantity must be multiplied by -1 for hole DQDs. The orange curve representing the triplets is dashed to help visualize the blue to purple transition of the $S(1,1)$ state. At the boundary between the (1,1) and (0,2) regions, the excited two-particle state $S(0,2)$ (purple curve)  anti-crosses and mixes with the ground state $S(1,1)$. Both states thus exchange their position before separating again at larger detuning. If the polarity of $V_{P1}$ and $V_{P2}$ would be reversed, i.e., for $\varepsilon_{V_P}<0$, the behavior of the $S(1,1)$ and $T(1,1)$ states would be mirrored, while $S(0,2)$ and $S(2,0)$ (not shown, at lower energy) would swap position. 

It should be noted that the behavior of the highest energy states (i.e; singly-unoccupied singlet and triplet) at $\varepsilon_{V_P}<$4~mV exhibits numerical noise in the $10^{-8}$~eV range. We attribute this phenomenon to the fact that none of the single-particle basis states is purely localized in one of the two dots when the detuning is close to 0~mV. As soon as $\varepsilon_{V_P}$ increases, the localization of the single-particle basis states becomes stronger, thus leading to more stable results. A similar behavior was observed by others when computing the exchange coupling of planar coupled hole spin qubits in Ge \cite{rodriguez2025dressed}. 
 
Notably, our CI calculations also return the eigenvectors corresponding to the five states of Fig.~\ref{fig:iso_eigenvectors}. They are expressed as linear combinations of Slater determinants according to Eq.~(\ref{eq:Phi}). This information enables us to visually highlight the localization and reordering of the states when the bias difference between the left and right plunger gates increases. In particular, we marked in Fig.~\ref{fig:iso_eigenvectors} three distinct detuning values by green geometrical forms. The isosurfaces of the associated eigenvectors are depicted in Fig.~\ref{fig:iso_eigenvectors} (bottom). Deep inside the (1,1) region (green square), the singly-unoccupied singlet, $S(1,1)$ (blue isosurface), and triplet, $T(1,1)$ (orange isosurface), have the highest energy and are delocalized over the two plunger gates, with a slightly enlarged charge on the right due to the positive detuning. In this regime, the doubly-unoccupied singlets are $S(2,0)$ and $S(0,2)$, which are represented as the brown and purple isosurfaces, respectively. They lay at lower energies, and only weakly interact with the higher-lying delocalized states. At the anti-crossing point (green triangle), the formerly right-localized $S(0,2)$ state (purple isosurface) starts spreading over both QDs. At the same time, $S(1,1)$ moves toward the right dot. While these states weakly interact at low $\varepsilon_{V_{P}}$, they strongly ``feel'' each other through Coulomb repulsion at $\varepsilon_{V_{P}}\sim$27~mV. As a consequence, band mixing occurs, giving rise to the hybridized $S_+$ (blue isosurface) and $S_-$ (purple isosurface) states. They correspond to the bonding and anti-bonding linear combination of $S(1,1)$ and $S(0,2)$ \cite{burkard_coupled_1999}. If the detuning further increases (green diamond), the mixing between $S(1,1)$ and $S(0,2)$ reduces and both states recover their localized/delocalized character, except that $S(0,2)$ (blue isosurface) is now the ground state, whereas the energy of $S(1,1)$ (purple isosurface) becomes slightly lower than that of the triplet state $T(1,1)$ (orange isosurface).     

\begin{figure}
\includegraphics[width=\columnwidth]{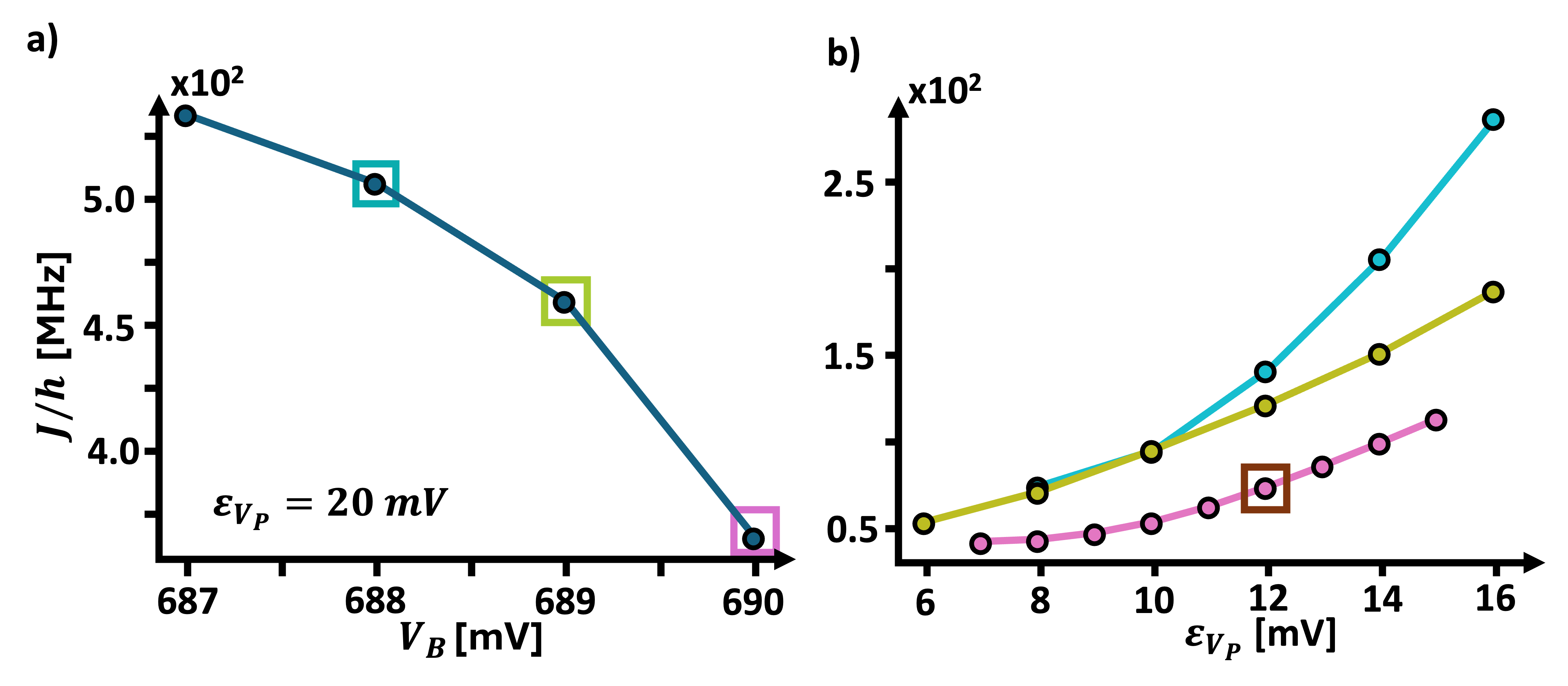}
\caption{\label{fig:device_biasing} (a) Exchange coupling $J/h$ in MHz as a function of an increasing $V_B$ bias at $V_{P1}/V_{P2}=494/474$~mV $\iff\varepsilon_{V_P}=$20~mV in the device from Fig.~\ref{fig:device}, with $h$ being Planck's constant. (b) Exchange coupling $J/h$ as a function of $\varepsilon_{V_P}$ for the values of $V_B$ marked by the colored squares in (a) for $V_{P1}(V_{P2})$ ranging from 487 (481)~mV to 492 (476)~mV.} 
\end{figure}

The electrostatic control of the DQD formed in our 5-gate FinFET is demonstrated in Fig.~\ref{fig:device_biasing} where we plot the tuning of the exchange coupling $J$ with respect to the barrier gate voltage $V_B$ and difference between the plunger gates $V_{P1}$ and $V_{P2}$. Note that in Fig.~\ref{fig:device_biasing} the exchange coupling is divided by the Planck's constant $h$, changing its unit from eV to Hz. In Fig.~\ref{fig:device_biasing}(a) $V_B$ is increased from 687~mV up to 690~mV at a constant plunger detuning $\varepsilon_{V_P}$=20~mV. By doing so, we make the potential barrier between the two dots higher and reduce both the overlap of their wavefunctions and the resulting exchange interaction $J$. Adjusting $V_B$ is indeed an efficient way to change the magnitude of $J$ as a small ramping of the barrier gate by 20~mV leads to a drop of the exchange coupling from 525 down to 350 MHz. In Fig.~\ref{fig:device_biasing}(b) we evaluate the exchange coupling $J$ at different $V_B$ values when the detuning $\varepsilon_{V_P}$ goes from 6~mV up to 16~mV. Increasing $\varepsilon_{V_P}$ further separates the singly-unoccupied singlet and triplet states, as depicted in the energy spectrum of Fig.~\ref{fig:iso_eigenvectors}, which enhances the exchange coupling. Compared to the $V_B$-induced modulation of $J$, the one obtained by tuning the plunger gates is more moderate and limited to a few tens of MHz only for the same detuning range. Overall, by combining $V_B$ and $V_P$ the exchange coupling in our two-qubit device can be precisely controlled over a large range.

\subsection{Two-qubit Quantum Logic Gates}

Careful engineering of the exchange interaction $J$ through external voltages and magnetic fields is key for the successful manipulation of coupled qubits and the realization of quantum logic gates. DQD architectures such as the one studied here can be leveraged to implement two-qubit gates such as $SWAP$ \cite{loss_quantum_1998,burkard_coupled_1999}, $CZ$ (or $CPHASE$) \cite{veldhorst2015two}, and $CNOT$ \cite{zajac2018resonantly}. While electric-dipole-spin-resonance (EDSR) induces single-qubit rotations on individual spin-qubits through AC resonant microwave at the plunger gate, the manipulation of a DQD system is achieved by taking advantage of the exchange interaction. More precisely, at high potential barrier ($V_B>720$~mV) the exchange between the two isolated holes can be switched on reducing the barrier height through DC pulsing of $V_B$ within the interaction range ($V_B<700$~mV). The exchange pulse duration, $\tau_J$, during which the coupled system interacts must be precisely controlled to obtain the desired quantum operation under an applied magnetic field \cite{tanttu2024assessment}.  

For $\varepsilon_{V_P}$ values lower than 15~mV (singlets' anticrossing point), in regions where $S(0,2)$ and $S(1,1)$ weakly interact, our two-qubit system can be effectively reduced to its four highest lying, singly-unoccupied singlet and triplet states $S(1,1)$ and $T(1,1)$ (triply degenerate), respectively. The application of a small magnetic field splits the degenerate triplet states, producing three individual triplet states, $T_+(1,1)$, $T_0(1,1)$, and $T_-(1,1)$. The magnetic response of our DQD system can be modeled according to the following effective Hamiltonian, which relies on the $S(0,2)$, $S(1,1)$, $T_+(1,1)$, $T_-(1,1)$, and $T_0(1,1)$ states as basis elements \cite{geyer2024anisotropic,saez2025exchange,stepanenko2012singlet}: 
\begin{eqnarray}
\label{eq:HB}
H =\begin{pmatrix}
 U-\varepsilon & \sqrt{2}t & 0 & 0 & 0 \\
 \sqrt{2}t & 0 & -\frac{\delta b_x + i\delta b_y}{\sqrt{2}} & \frac{\delta b_x - i\delta b_y}{\sqrt{2}} & \delta b_z \\
0 & -\frac{\delta b_x - i\delta b_y}{\sqrt{2}} & \bar{b}_z & 0 & \frac{\bar{b}_x - i\bar{b}_y}{\sqrt{2}} \\
0 & \frac{\delta b_x + i\delta b_y}{\sqrt{2}} & 0 & -\bar{b}_z & \frac{\bar{b}_x + i\bar{b}_y}{\sqrt{2}} \\
0 & \delta b_z & \frac{\bar{b}_x + i\bar{b}_y}{\sqrt{2}} & \frac{\bar{b}_x - i\bar{b}_y}{\sqrt{2}} & 0
\end{pmatrix}.
\end{eqnarray}
Here, $\boldsymbol{\bar{b}}=\mu_B(R(-\theta_{so})\underline{g_L}+R(\theta_{so})\underline{g_R})\boldsymbol{B}/2$ and $\boldsymbol{\delta b}=\mu_B(R(-\theta_{so})\underline{g_L}-R(\theta_{so})\underline{g_R})\boldsymbol{B}/2$, where $\underline{g_{L/R}}$ is the symmetric $g$-tensor of the right/left dot. $R$ is a rotation matrix around the SOI axis with unit vector $\boldsymbol{\hat{n}}_{so}$ by an angle $2\theta_{so}\approx 2d/\lambda_{so}$, with $d$ being the distance between the dots and $\lambda_{so}$ being the spin-orbit length. Finally, $\mu_B$ is the Bohr magneton, and $\boldsymbol{B}$ is the magnetic field vector. 

Inspecting the Hamiltonian in Eq.~(\ref{eq:HB}) reveals that the off-diagonal block coupling the singlet state $S(1,1$ to the triplet states depends on the difference in Zeeman energy, $\delta E_Z = \mu_B|(R(-\theta_{so})\underline{g_L}-R(\theta_{so})\underline{g_R})\boldsymbol{B}|/2$. Hence, two operating regimes can be identified: an exchange-dominated regime characterized by $\delta E_Z \ll J$, and a Zeeman-dominated regime where $\delta E_Z \gg J$. Both regimes can be taken advantage of to create specific two-qubit quantum logic gates, namely a SWAP gate (exchange-dominant) and a Controlled-Phase or CZ gate (Zeeman-dominant) \cite{dots2005coherent, meunier2011efficient, russ2018high, geyer2024anisotropic, brunner2011two, asai2023device}. In the standard computational basis for two qubits $\{|00\rangle, |01\rangle, |10\rangle, |11\rangle\}$, the SWAP gate exchanges their states, whereas the CZ gate applies a $-1$ phase exclusively to the $|11\rangle$ state.

In systems with strong spin-orbit coupling, the exchange interaction is anisotropic, splitting into transverse ($J_\perp$) and longitudinal ($J_\parallel$) components within the QD frame \cite{geyer2024anisotropic}. In the exchange-dominated regime, the transverse component $J_\perp$ drives ``flip-flop'' transitions. A prepared product state $|\Uparrow\Downarrow\rangle$, which is not an eigenstate of the system, will coherently evolve into $|\Downarrow\Uparrow\rangle$ over a time $\tau_{\text{SWAP}} = h / J_\perp$ \cite{loss_quantum_1998}, effectively swapping the spin states. Interrupting this pulse halfway at $\tau = h / (2J_\perp)$ realizes a maximally entangling $\sqrt{\text{SWAP}}$ gate.  

Conversely, in the Zeeman-dominated regime ($\delta E_Z \gg J_\perp$), flip-flop transitions are energetically suppressed, and the system eigenstates are well-approximated by the separable states $|\Uparrow\Downarrow\rangle$ and $|\Downarrow\Uparrow\rangle$, allowing for Pauli spin blockade readout \cite{hanson2007spins, jirovec2021singlet}. In this regime, the longitudinal exchange component $J_\parallel$ provides an Ising-like interaction. This interaction shifts the energy of the anti-parallel states ($|\Uparrow\Downarrow\rangle, |\Downarrow\Uparrow\rangle$) relative to the parallel states ($|\Uparrow\Uparrow\rangle, |\Downarrow\Downarrow\rangle$) by $J_\parallel$. By applying this exchange interaction for a duration $\tau_{\text{CZ}} = h / (2J_\parallel)$, the parallel and anti-parallel manifolds accumulate a relative phase of $\pi$. This realizes a two-qubit controlled-phase gate, which is locally equivalent to a standard CZ gate.\cite{meunier2011efficient,veldhorst2014addressable}.

\begin{figure*}
\includegraphics[width=\textwidth]{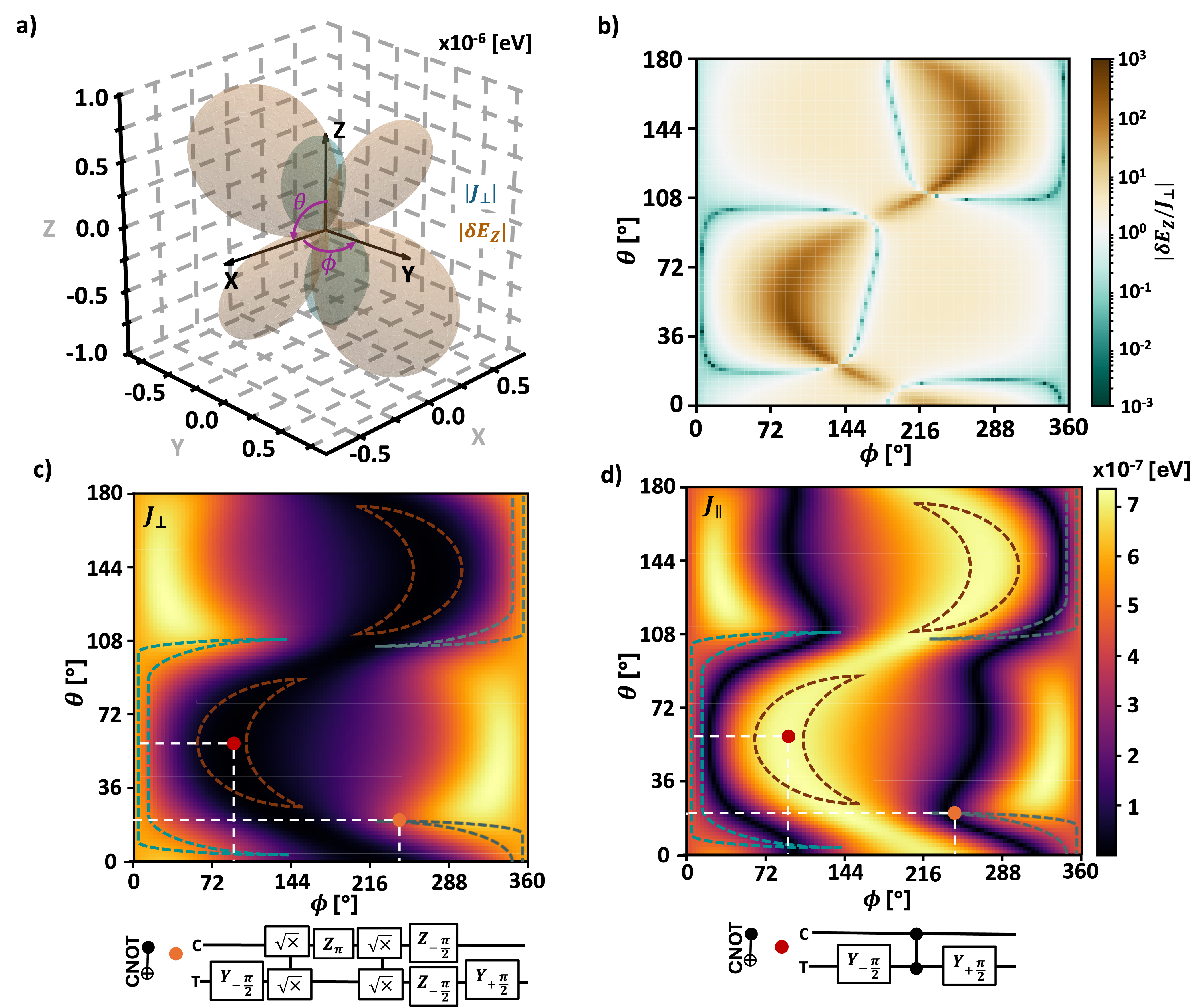}
\caption{\label{fig:magnetic_plots} (a) Three-dimensional angular dependence of the Zeeman energy difference $|\delta E_Z|$ (brown) and the transverse exchange magnitude $|J_\perp|$ (turquoise) for the simulated DQD. (b) Logarithmic colormap illustrating the ratio between the Zeeman energy difference and the transverse exchange. Dark brown regions favor Zeeman-dominated CZ operations, while dark turquoise regions favor exchange-dominated SWAP operations. (c) Contour map of the transverse exchange component $J_\perp$ as a function of the magnetic field angles $\theta$ and $\phi$. Black regions denote areas where the flip-flop mechanism is suppressed, while bright yellow regions indicate maximum exchange strength. The orange dot marks a highly favorable orientation for implementing a SWAP gate. (d) Contour map of the longitudinal exchange component $J_\parallel$. The red dot highlights an optimal ``sweet spot'' for high-fidelity CZ gates, where the required $J_\parallel$ is finite but the transverse exchange $J_\perp$ vanishes, completely suppressing leakage errors. (Bottom) Circuit schematics showing the realization of a two-qubit CNOT gate from (left) two native $\sqrt{\text{SWAP}}$ operations combined with physical $Y$ and virtual $Z$ rotations, and (right) a single native CZ gate flanked by target qubit $Y_{\pm\pi/2}$ rotations. The letters C and T stand for ``control'' and ``target'' qubit, respectively.} 
\end{figure*}

To demonstrate the suitability of our 5-gate FinFET as two-qubit gates, we now assess the aforementioned interactions inside the reference structure from Fig.~\ref{fig:device}. The required $g$-tensors $\underline{g_{L/R}}$ in Eq.~(\ref{eq:HB}) are constructed from the $g$-matrix \cite{venitucci2018electrical} of the spin-mixed single-particle bounding and antibounding states obtained from the SP simulation in step (a) of Fig.~\ref{fig:methodology}. At zero detuning, a good estimate of the tunnel coupling is given by $t=\Delta/\sqrt{2}$, where $\Delta$ is the energy separation between the bonding and anti-bonding state as shown in Fig.~\ref{fig:bandmixing}(a) (i.e., $t$=1.13e-5~eV. Moreover, in the regime where $S(0,2)$ and $S(1,1)$ weakly interact, the exchange coupling can be approximated by the simplified expression $J=\frac{2t^2}{U-\varepsilon}$ \cite{geyer2024anisotropic,qtcad}, where $U$ is the Coulomb interaction. The term $U-\varepsilon$ can be evaluated using the $J$ obtained from the Configuration Interaction (CI) routine at the desired detuning. The remaining quantities, $\lambda_{so}$ and $\boldsymbol{\hat{n}}_{so}$, were taken from the experimental measurements in Ref.~\cite{geyer2024anisotropic}. Following the methodology outlined in the latter reference the exchange coupling matrix for the anisotropic exchange can be reconstructed as $\hat{J}=JR(-2\theta_{so})$. To extract $J_{\perp}$ and $J_{\parallel}$, we move to the  qubit frame applying the following rotation: 
\begin{eqnarray}
\label{eq:J_QD}
\tilde{J}=R_L\hat{J}R_R^T,
\end{eqnarray}
the matrices $R_L$ and $R_R$ are the mathematical transformations required to rotate the laboratory frame's $z$-axis $(0,0,1)$ so that it points exactly along $\vec{n}_L$ and $\vec{n}_R$, the local quantization axes for each qubit is given by the vector $$\vec{n}_{L/R} = \frac{\underline{g_{L/R}} \boldsymbol{B}}{|\underline{g_{L/R}} \boldsymbol{B}|}$$. We then define the perpendicular and parallel exchange components as:
\begin{subequations}
\label{eq:J_anis}
\begin{eqnarray}
\label{J_per}
|J_\perp| = \frac{1}{2} \sqrt{ (\tilde{J}_{xx} + \tilde{J}_{yy})^2 + (\tilde{J}_{xy} - \tilde{J}_{yx})^2 },
\end{eqnarray}
\begin{eqnarray}
\label{eq:J_par}
J_{\parallel}=\tilde{J}_{zz}.
\end{eqnarray}
\end{subequations}
In Fig.~\ref{fig:magnetic_plots}(a), we plot the variation of $|\delta E_Z|$ (in brown) and $|J_\perp|$ (in turquoise) with respect to the magnetic field orientation for the simulated DQD system. This three-dimensional representation outlines the most favorable areas for the implementation of our selected two-qubit gates. It appears that aligning the magnetic field strictly along the device axes does not provide sufficient asymmetry between the holes ($\delta E_Z \sim 0$). A stronger disparity between the left and right dot is achieved for magnetic fields directed outside the high-symmetry planes. This is clearly evidenced in Fig.~\ref{fig:magnetic_plots}(b), where the absolute ratio $|\delta E_Z / J_\perp|$ is displayed on a logarithmic scale. Dark brown areas, such as the maximum located near $(\theta,\phi)\approx(40^\circ,100^\circ)$, correspond to magnetic field orientations where $\delta E_Z$ dominates, favoring the implementation of a CZ gate. Conversely, the dark turquoise regions denote the dominant $J_\perp$ required for efficient SWAP operations.  

Using the equations defined above, we present the values of $J_\perp$ and $J_\parallel$ in Fig.~\ref{fig:magnetic_plots}(c) and Fig.~\ref{fig:magnetic_plots}(d), respectively, as a function of the azimuthal angle $\phi$ and the polar angle $\theta$. The ability to navigate this angular landscape is critical for high-fidelity quantum operations. To implement a high-fidelity CZ gate, it is highly desirable to operate at a magnetic field orientation where the transverse exchange is entirely suppressed ($J_\perp = 0$). In this regime, represented by the red dot in Fig.~\ref{fig:magnetic_plots}(c) and (d), the Ising-like $J_\parallel$ interaction can drive the conditional phase shift without the risk of unwanted flip-flop transitions leaking information out of the computational basis. On the other hand, the implementation of a SWAP gate relies explicitly on $J_\perp$ and is remarkably robust against the presence of $J_\parallel$. Because $J_\parallel$ shifts the energy of the anti-parallel states $|\Uparrow\Downarrow\rangle$ and $|\Downarrow\Uparrow\rangle$ by the exact same amount, their energy degeneracy is preserved, allowing the $J_\perp$-driven flip-flop mechanism to proceed unimpeded (orange dot in Fig.~\ref{fig:magnetic_plots}). The only consequence of a non-zero $J_\parallel$ during a SWAP operation is the accumulation of an additional phase on the parallel states. However, this  ``phase tax'' does not degrade the gate fidelity, as it can be perfectly compensated for via virtual $Z$-gate corrections applied in software immediately following the two-qubit pulse \cite{mckay2017efficient,zajac2018resonantly,mills2022two}.

Ultimately, distinct magnetic orientations enable native $\sqrt{\text{SWAP}}$ or CZ gates, which form a universal gate set when combined with AC-driven single-qubit rotations. The bottom panel of Figure~\ref{fig:magnetic_plots} shows the CNOT circuit decompositions utilizing both (i) $\sqrt{\text{SWAP}}$ and (ii) CZ blocks. Based on previous simulations \cite{bouquet2025simulation}, the physical $Y_{\pm\pi/2}$ rotations were estimated at approximately $25$~ns (10 MHz Rabi frequency) at $B=50$~mT, while $Z$-rotations can be implemented purely as virtual software phase shifts ($0$~ns). Disregarding the physical implementation overhead and focusing only on the manipulation time, the operational speed is determined solely by the two-qubit interaction.

\subsection{Comparison with experimental quantum device}

\begin{figure}
\includegraphics[width=\columnwidth]{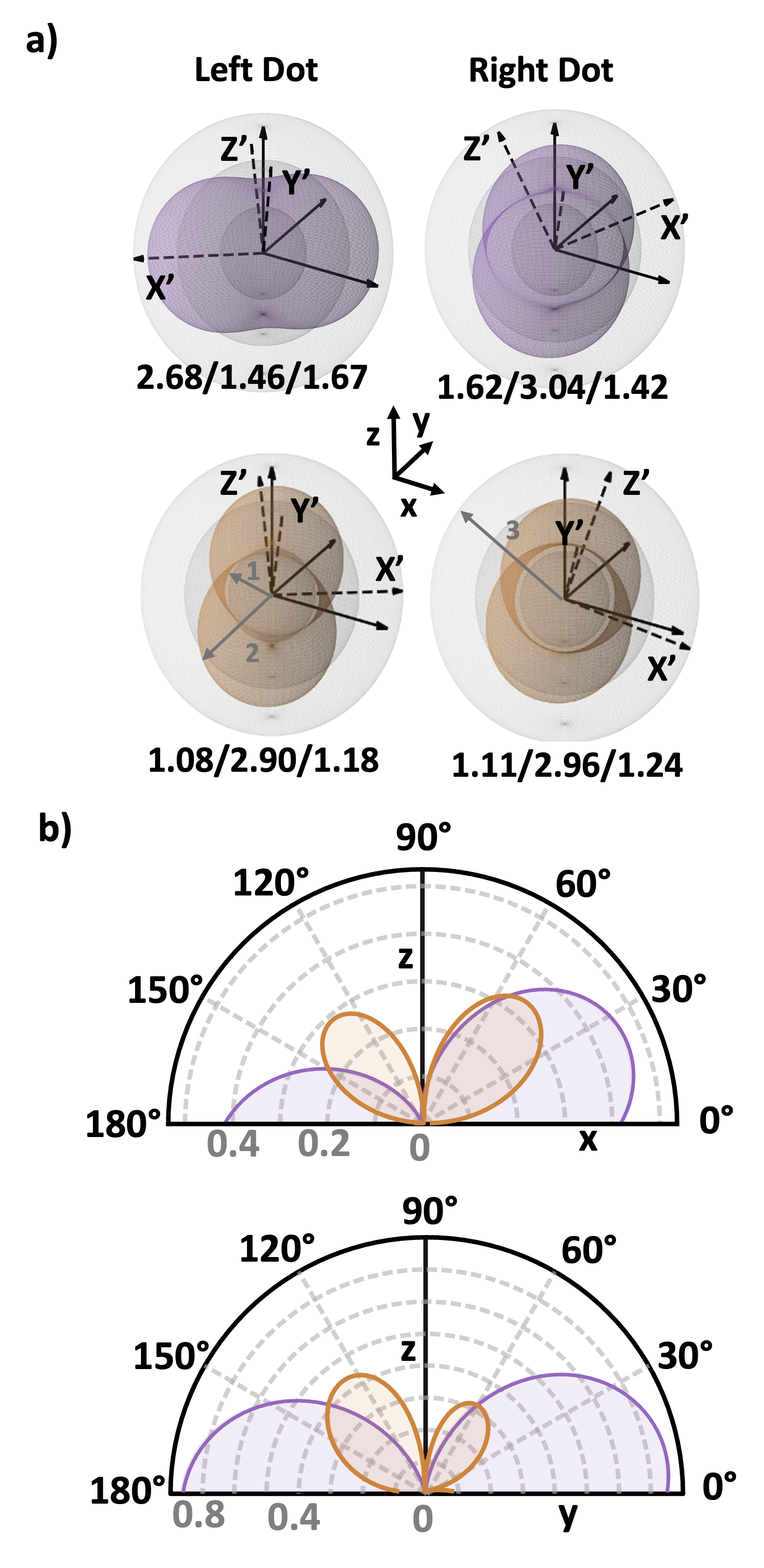}
\caption{\label{comparison_g} Comparison of the magnetic response between the experimental device from \cite{geyer2024anisotropic} (purple) and the simulated structure of Fig.~\ref{fig:device} (brown) under the bias conditions marked by the brown square in Fig.~\ref{fig:device_biasing}(b). (a) Representation of the $g$-tensor corresponding to the left and right quantum dot for the experimental (top) and simulated device (bottom). The $g$-factors along the three principal magnetic directions are given as $g_{X'}/g_{Y'}/g_{Z'}$. The solid arrows indicate the device axes while the dashed arrows refer to the magnetic axes of the $g$-tensors. The cartesian coordinates of the magnetic axes are given in Appendix~\ref{A:magnetic_axes}. (b) Polar plot of the $\delta E_Z$ per unit of Tesla in the $xz$-(top) and $yz$-(bottom) plane for both devices (purple: experiments, brown: simulations)}
\end{figure}

We now examine the magnetic response of our simulated geometry from Fig.~\ref{fig:device} and compare our results to measurements of the experimental device from Ref.~\cite{geyer2024anisotropic}. First, the simulated and experimental $g$-tensors of the left and right dots are plotted in Fig.~\ref{comparison_g}. Although the finite difference scheme implemented in our SP solver does not allow us to capture the exact contour of the different device material layers, symmetries can still be enforced with high accuracy. As a result, the simulated QDs display almost identical $g$-tensors, though with different magnetic orientations. In contrast, the experimental data reveals a strong tilt between the left and right QDs, displaying $g$-factors almost orthogonal to each other. Although, practically, two dots can probably not be made perfectly identical, it is still expected that their $g$-tensors remain very similar as these quantities primarily depend on the channel geometry and gate electrode architecture. While the experimental right dot shows remarkably close orientation and magnitude with the simulated $g$-tensors, this is however not the case for the left one. Its measured $g_{X'}$/$g_{Y'}$/$g_{Z'}$ orientation suggests the presence of geometrical inhomogeneities that might play a critical role. For example, interfacial charges and/or surface roughness could effectively anchor the wavefunction of the left QD and lead to the observed fluctuations in the measurements \cite{martinez_variability_2022}. Also, through transmission electron microscopy images, variations of a few nanometers between supposedly identical plunger and barrier gate widths have been identified, providing another explanation for the $g$-tensor asymmetries. As a consequence, large $\delta E_Z$'s may be artificially induced for specific magnetic field orientations. This effect is highlighted in Fig.~\ref{comparison_g}(b) where $\delta E_Z/\mu_B$ is plotted within the device's high-symmetry planes. The difference between both dots can be clearly noticed, especially in the $xz$-plane (top), the right lob being more prominent than the left one in the experiments, while they are symmetric in the simulations. It is therefore difficult to compare the simulated and measured exchange coupling $J$, which is extracted from the $S-T_0$ energy separation, as it strongly depends on $\delta E_Z/\mu_B$ through Eq.~(\ref{eq:HB}). As expected, the agreement between both datasets is limited quantitatively, although they exhibit a similar qualitative trend, namely a central dip at $\alpha\sim90^\circ$ and two asymmetric peaks located near the left and right plot edges.

\section{Conclusion}\label{sec:Conclusion}

The exchange interaction arising from the coupling of adjacent quantum dots represents the cornerstone for the realization of two-qubit quantum logic gates. Through quantum mechanical investigations based on self-consistent Schrödinger-Poisson simulations, we have determined this parameter in a 5-gate, DQD FinFET-like structure with a triangular cross section and inspired from an experimental device. Starting from single-particle wavefunctions, we have constructed a configuration interaction Hamiltonian that returns the energy of singlet and triplet states as a function of the detuning between the two plunger gates. Next, the influence of external magnetic fields has been added through a dedicated Hamiltonian whose diagonalization returns the exchange coupling response of the simulated structure with respect to the field orientation. Our simulations have revealed that the investigated 5-gate FinFET geometry features magnetic sweet-spots where either a $SWAP$ gate (pulsable to $\sqrt{SWAP}$) or a $CZ$ gate can be implemented. Combined with AC-driven EDSR for single-qubit rotations, these sweet-spots offer two distinct pathways to realize a universal quantum gate set from which any mathematically possible quantum computation can be performed. This flexibility highlights the potential of Si FinFET-hosted hole spin qubits as scalable building blocks of quantum processors. Finally, our simulated exchange coupling has been compared to experiments, but due to unexpected asymmetries in the measured $g$-tensors, the agreement between both datasets is sub-optimal. 

As reducing intra-device and device-to-device variability is a challenging process, improving the quality of the comparison with experiments would require simulating a large number of structures with random variations of their main geometrical parameters. The results should be either statistically averaged or combined to identify the configuration with the highest resemblance to the experimental reference. Moreover, since charge noise remains a limiting factor for silicon-based spin qubits, future work will focus on incorporating this effect into our simulation platform. For example, different crystallographic orientations could be investigated to identify the one exhibiting the least amount of charge noise. The ability to determine possible sweet-spots \textit{in silico} is of high relevance to guide the design of future hole spin qubits and to substantially enhance their coherence time and gate fidelity.

\section*{Acknowledgment}
This work was supported by the Swiss National Science Foundation under the NCCR SPIN (grant $\mathrm{n^\circ}$ 225153) and the QuaTrEx project (grant $\mathrm{n^\circ}$ 209358). The usage of CSCS computing resources under project lp82 is acknowledged.

\appendix
\section{\label{A:parameters}Simulations parameters for the self-consistent Schr\"odinger-Poisson simulations}
\begin{table}[h]
\caption{\label{tab:table1}%
List of materials parameters used in the thermal and SP simulations: Young's modulus ($E$), thermal expansion coefficient ($\alpha$), Poisson's coefficient ($\nu$), Luttinger's parameters ($\gamma_{1}$, $\gamma_{2}$, and $\gamma_{3}$), split-off spin-orbit energy ($\Delta_{0}$), isotropic magnetic parameter ($\kappa$), and deformation potentials ($a_{\nu}$, $b$, and $d$) \cite{winkler_spin-orbit_2003}. }
\begin{ruledtabular}
\begin{tabular}{cccccccc}
&$E$&$\alpha\times10^{-6}$&$\nu$&$\gamma_{1}$&$\Delta_{0}$&$\kappa$&$a_{\nu}$ \\
&[GPa]&[$K^{-1}$].&&$\gamma_{2}$&[eV]&&$b$ \\ 
&&&&$\gamma_{3}$ &&&$d$ \\ 
\hline
Si&169&2.6&0.27&4.285&0.044&-0.42&2.46\\
&&&&0.339&&&-2.35\\
&&&&1.21&&&-5.32\\
SiO$_{2}$&73&0.49&0.17&-&-&-&-\\
TiN&43&9.35&0.33&-&-&-&-
\end{tabular}
\end{ruledtabular}
\end{table}

\section{\label{A:magnetic_axes}Magnetic axes of the experimental and simulated hole-qubit pair.}
\begin{table}[h]
\caption{\label{tab:table2}%
Cartesian coordinates of the magnetic axes generated from the diagonalization of the $g$-matrices corresponding to the $g$-tensor representations given in Fig.~\ref{comparison_g}(a) for the experimental (Exp.) and simulated (Sim.) device.}
\begin{ruledtabular}
\begin{tabular}{ccccccc}
&&Exp. Left Dot&&&Exp. Right Dot \\
\hline
&X'&Y'&Z'&X'&Y'&Z' \\
\hline
x&-0.81&0.27&-0.53&0.81&-0.44&0.39\\
y&-0.59&-0.35&0.71&0.41&0.90&-0.16\\
z&0.01&0.90&0.44&-0.03&0.90&0.43\\
\end{tabular}
\end{ruledtabular}
\end{table}

\begin{table}
\begin{ruledtabular}
\begin{tabular}{ccccccc}
&&Sim. Left Dot&&&Sim. Right Dot& \\
\hline
&X'&Y'&Z'&X'&Y'&Z'\\
\hline
x&0.91&-0.42&-0.02&0.87&-0.41&0.26\\
y&0.42&0.90&-0.16&0.36&0.91&0.22\\
z&0.08&0.14&0.98&-0.32&0.10&0.94\\
\end{tabular}
\end{ruledtabular}
\end{table}

\bibliography{bibliography}

\end{document}